\documentclass[conference]{IEEEtran}
\IEEEoverridecommandlockouts
\usepackage{cite}
\usepackage{amsmath,amssymb,amsfonts}
\usepackage{algorithm}
\usepackage{algorithmic}
\usepackage{graphicx}
\usepackage{textcomp}
\usepackage{xcolor}
\usepackage{listings}
\usepackage{tabularx}
\usepackage{siunitx}
\usepackage{booktabs}
\usepackage{caption}
\usepackage{comment}
\usepackage{balance}
\usepackage{array}
\usepackage{balance}
\usepackage{subcaption}

\usepackage{acro}
\DeclareAcronym{MBE}{
	short = MBE,
	long  = Model-based Engineering,
	sort  = abbrev,
}

\DeclareAcronym{LLM}{
	short = LLM,
	long  = Large Language Model,
	sort  = abbrev,
}

\DeclareAcronym{XMI}{
	short = XMI,
	long  = XML Metadata Interchange,
	sort  = abbrev,
}

\DeclareAcronym{XML}{
	short = XML,
	long  = Extensible Markup Language,
	sort  = abbrev,
}

\DeclareAcronym{OMG}{
	short = OMG,
	long  = Object Management Group,
	sort  = abbrev,
}

\DeclareAcronym{MOF}{
	short = MOF,
	long  = Meta-Object Facility,
	sort  = abbrev,
}

\DeclareAcronym{EMF}{
	short = EMF,
	long  = Eclipse Modeling Framework,
	sort  = abbrev,
}

\DeclareAcronym{UML}{
	short = UML,
	long  = Unified Modeling Language,
	sort  = abbrev,
}

\DeclareAcronym{SysML}{
	short = SysML,
	long  = Systems Modeling Language,
	sort  = abbrev,
}

\DeclareAcronym{DSL}{
	short = DSL,
	long  = Domain Specific Language,
	sort  = abbrev,
}

\begin{document}

\title{LLM-enabled Instance Model Generation}

\author{

	\author{
			Fengjunjie Pan, Nenad Petrovic, Vahid Zolfaghari, Long Wen, and Alois Knoll}
		\thanks{F. Pan, N. Petrovic, V. Zolfaghari,  L. Wen, A. Knoll are with Robotics, Artificial Intelligence and Real-Time Systems, School of Computation, Information and Technology, Technical University of Munich, Munich, Germany. \{panf, pne, zov, wenl, knoll\}@in.tum.de}%

}

\maketitle

\begin{abstract}
In the domain of model-based engineering, models are essential components that enable system design and analysis. 
Traditionally, the creation of these models has been a manual process requiring not only deep modeling expertise but also substantial domain knowledge of target systems.
With the rapid advancement of generative artificial intelligence, large language models (LLMs) show potential for automating model generation.
This work explores the generation of instance models using LLMs, focusing specifically on producing XMI-based instance models from Ecore metamodels and natural language specifications.
We observe that current LLMs struggle to directly generate valid XMI models. To address this, we propose a two-step approach: first, using LLMs to produce a simplified structured output containing all necessary instance model information, namely a conceptual instance model, and then compiling this intermediate representation into a valid XMI file. 
The conceptual instance model is format-independent, allowing it to be transformed into various modeling formats via different compilers.
The feasibility of the proposed method has been demonstrated using several LLMs, including GPT-4o, o1-preview, Llama 3.1 (8B and 70B).
Results show that the proposed method significantly improves the usability of LLMs for instance model generation tasks. Notably, the smaller open-source model, Llama 3.1 70B, demonstrated performance comparable to proprietary GPT models within the proposed framework.

\end{abstract}


\section{Introduction}
\ac{MBE} has been widely discussed in designing complex systems in industries such as aviation~\cite{Li2019}, automotive~\cite{DAmbrosio2017} and robotics~\cite{Wen2025}.
It provides a structured methodology to represent systems and leverages formal methods for their analysis.
The \ac{MOF}, defined by \ac{OMG} \cite{OMG2016}, proposes a standardized framework for \ac{MBE}.
It introduces a layered modeling architecture where every element in a model corresponds to an element at a higher abstraction level.
Common modeling languages, such as \ac{UML}~\cite{OMG2017} and \ac{SysML}~\cite{OMG2024}, conform to this standard.
In modeling tasks, a metamodel is an abstraction that defines the structure, components, and rules of a system. A concrete system is presented by an instance model that conforms to its corresponding metamodel. Creating instance models typically requires both proficiency in modeling languages and a deep understanding of the abstract system information encoded in the metamodel. This process can be time-consuming and error-prone, especially when performed manually.
Given the demonstrated capabilities of \acp{LLM}  in generative tasks such as code generation, we propose an \ac{LLM}-based approach for the automated instance model generation.

\begin{figure}[t]
	\centering
	\includegraphics[width=\linewidth]{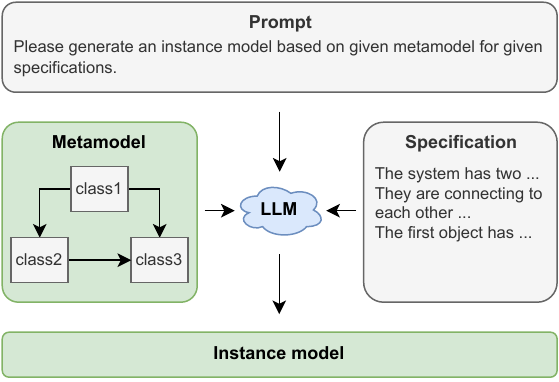}
	\caption{Generation of an instance model based on a given metamodel and specifications. The prompt and specifications are provided as textual input to the LLM The metamodel and instance model are exchanged in a modeling format.}
	\label{fig:method}
\end{figure}

In this work, we investigate the automated creation of instance models from given metamodels and textual system specifications (Fig.~\ref{fig:method}). 
We mainly focus on models compatible with the \ac{EMF}~\cite{Steinberg2008}, a widely used tool based on the MOF standard for creating and managing both metamodels and instance models. 
\ac{EMF} uses the Ecore format to present meta model, and employs \ac{XMI} \cite{OMG2015} for saving instance models.
XMI, standardized by OMG, provides a common format for storing and exchanging  model information.
We notice that current LLMs are facing challenges to directly generate valid XMI instance models due to nested structure and verbose syntax in the model format.
To address this challenge, we introduce a simplified intermediate format called the conceptual instance model. This textual representation retains all required model information but avoids the syntactic complexity of XMI.
By offloading the XMI complexity, our approach allows LLMs to focus on generating semantically meaningful content.
To complete the process, we develop a transformation algorithm that converts conceptual instance models into syntactically valid XMI files.
It is worth noting that, although we use EMF-based models to illustrate the core principles of our method, the proposed approach is equally applicable to other MOF-compliant modeling languages, such as UML and SysML.

To evaluate the effectiveness of the proposed method, we constructed a test set consisting of Ecore metamodels with real-world semantics. For each metamodel, we generated human-reviewed natural language specifications and manually created reference instance models. We then conducted experiments using several state-of-the-art LLMs, including commercial models (GPT-4o and o1-preview) and open-source models (Llama 3.1-8B and 70B). The results demonstrate that our method significantly enhances the performance of instance model generation tasks. Furthermore, the findings indicate that smaller open-source models, such as Llama 3.1-70B, can achieve performance comparable to that of commercial GPT models within the proposed framework.


\section{Background and Related Work} \label{sec:background}
\subsection{Metamodeling and Instantiation}
\begin{figure}[t]
	\centering
	\includegraphics[width=\linewidth]{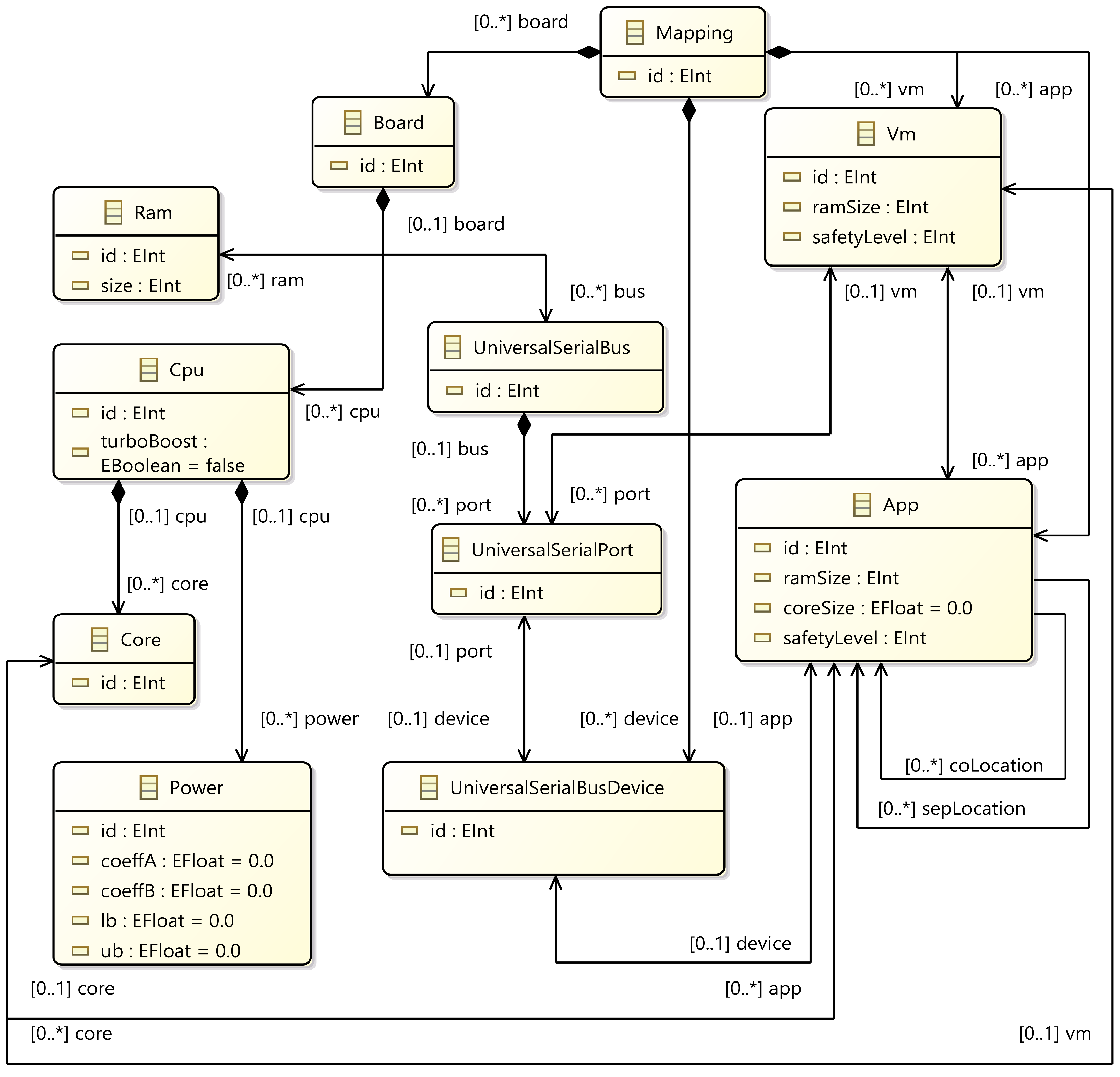}
	\caption{An example of a metamodel illustrating a virtualization-based software-hardware resource allocation system~\cite{Pan2023}. The metamodel is visualized using Eclipse EMF.}
	\label{fig:metamodel}
\end{figure}

Models are fundamental components in \ac{MBE}. 
The concept of metamodeling refers to the construction of abstract or meta-level representation of systems within a specific domain.
A metamodel defines the foundational building blocks of a system by specifying system components as classes, their properties as attributes, and the possible relationships or interactions among them as associations.
In this context, a metamodel serves as the abstract syntax for instance models.
An instance model is a concrete instantiation of a metamodel, representing a specific system. It consists of instance objects created based on the class definitions in the metamodel, assigns concrete values to attributes and establishing associations among objects.

Models can be represented in various forms, such as visual diagrams (e.g., class diagrams) or structured model files in different formats. 
In this work, we focus primarily on Ecore metamodels and XMI instance models.
These models can be loaded and processed by tools such as the Eclipse \ac{EMF}.
We consider the generation of graphical representations of models to be one of the processes supported by such tools. Therefore, the generation of graphical models using multimodal \acp{LLM} is beyond the scope of this work.

\begin{figure}[t]
	\centering
	\includegraphics[width=0.8\linewidth]{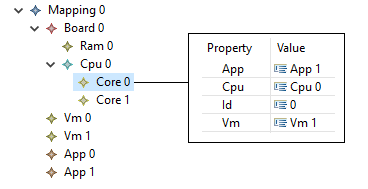}
	\caption{An instance model example created from the metamodel in Fig.~\ref{fig:metamodel}. This model was generated using the proposed method with GPT-4o. This instance model is visualized using Eclipse EMF.}
	\label{fig:insmodel}
\end{figure}

An example metamodel describing a software-hardware resource allocation system~\cite{Pan2023} is shown in Fig.~\ref{fig:metamodel}. 
In the illustration, yellow blocks represent classes, from which instances can be created. Each class defines a set of attributes, and associations between classes are represented by connecting lines: containment associations are indicated by lines with diamonds and arrows, while reference associations use simple arrows.
This metamodel outlines a general resource mapping system involving boards, virtual machines (VMs), and applications (APPs). A board may contain various subcomponents such as central processing units (CPUs) and random-access memory (RAM). These components can be further decomposed—for example, a CPU may consist of multiple cores.
The resource allocation scenario defined in the metamodel specifies that cores are assigned to VMs and will host APPs.
Based on this metamodel, a concrete instance model representing a specific resource allocation scenario is shown in Fig.~\ref{fig:insmodel}. This instance includes one Board with one RAM module and one CPU, two VMs, and two APPs. As shown, core0 is assigned to VM1, which hosts app1.
Both the metamodel and the instance model are visualized using Eclipse EMF to improve graphical readability. These visual representations are commonly used by engineers for manual analysis and documentation. The underlying model files are stored in Ecore or XMI format.

\subsection{Related Work}
Traditional methods for instance model generation have shown effectiveness in specific domains and use cases.
Ehrig et al.~\cite{Ehrig2009} proposed deriving a graph grammar from a metamodel to obtain an operational description of the system described in the metamodel.
This grammar is then used to systematically generate instance models conforming to the metamodel’s structure.
The focus of this study is primarily on generating class structures and basic associations, especially for those used in statechart models.
Hoffmann et al.~\cite{Hoffmann2011} extended the graph grammar approach to better distinguish between containment and reference associations. 
However, these methods do not explicitly address the generation of attribute values.

In addition to graph grammar techniques, solver-based approaches have also been discussed for instance model generation.
Torlak~\cite{Torlak2009} used relational logic to describe models, which is then translated into Boolean satisfiability~(SAT) problems and solved using an SAT solver. The resulting solutions can be then transformed back into model instances.
González et al.~\cite{González2012} translated EMF models into constraint satisfaction problems (CSPs) and used CSP solvers to generate valid model instances. 
While solver-based methods provide strong guarantees for constraint satisfaction, they naturally struggle to handle attributes with non-numeric type, such as strings, since solvers are typically designed for numerical and logical domains.

Nevertheless, all methods mentioned above primarily focus on the random generation of valid instance models, often intended for tasks like model transformation testing. In contrast, typical modeling tasks require that generated instance models also reflect specific semantics from given specifications.
With the recent advance of generative AI, researchers have begun investigating the use of LLMs in model generation tasks.

Petrovic et al.~\cite{Petrovic2025}  proposed an LLM-based approach for creating metamodels in the automotive domain. In their approach, requirements were iteratively provided as input to incrementally build the complete metamodel in Ecore format. 
C\'{a}mara et al.~\cite{Camara2023} evaluated the LLM in generating textual \ac{UML} class diagrams, noting that EMF Ecore model generation exhibits similar behavior to UML. They observed that while LLMs have shown promise, their performance in early modeling tasks remains limited due to frequent syntactic and semantic errors, especially when compared to code generation.
Ferrari et al.~\cite{Ferrari2024} explored the ability of ChatGPT for generating \ac{UML} sequence diagrams from natural language requirements. Their approach involved generating PlantUML~\cite{PlantUML} descriptions through carefully designed prompts and using external tools for diagram rendering. Their experiments show that challenges exist for \acp{LLM} in terms of model completeness and correctness.
While recent work has provided valuable insights into using LLMs for various types of model generation, no systematic analysis has been conducted for the general instance model generation topic. This task requires considering multiple dimensions at the same time, including the syntax defined by the metamodel, the grammar of the modeling language, and the semantic information described in the natural language specification.

In this work, we introduce a novel LLM-based approach for instance model generation. We introduce an intermediate representation, called the conceptual instance model, which offloads modeling-format grammar concerns from the LLM to a dedicated compiler program.
This separation enables the LLM to focus on extracting content from natural language specifications and aligning it with the structure defined in the metamodel, while the compiler ensures the validity of the target modeling format.
We compare our approach against direct generation baselines and demonstrate that our framework significantly improves both the usability of LLMs for instance model generation.

\section{Proposed Approach} \label{sec:proposed approach}
\begin{figure}[t]
	\centering
	\includegraphics[width=0.9\linewidth]{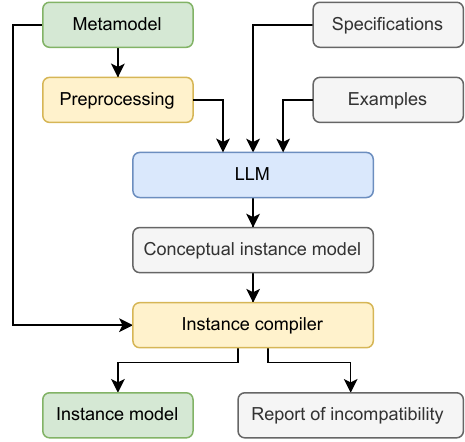}
	\caption{ Overview of the proposed method for instance model generation. Model files are shown in green, natural language-based texts in gray, and computer programs in yellow.}
	\label{fig:approach}
\end{figure}

We propose an \ac{LLM}-based approach for the generation of instance models based on metamodel information and natural language specifications (Fig.~\ref{fig:approach}). 
The workflow begins with the metamodel preprocessing. 
Metamodels are typically presented as \ac{XML}-based files, e.g., Ecore files, which can be relatively long and verbose due to extensive tagging and hierarchical structures. 
To simplify and structure this input for \acp{LLM}, we adopt PlantUML, a textual modeling language that has proven effective for both input and output representations in \ac{LLM}-based modeling tasks~\cite{Pan2024, Ferrari2024}.
 
The LLM takes as input both the processed metamodel information (in PlantUML format) and the system specification written in natural language.
Despite the strong capabilities of \acp{LLM} in general code generation, they often struggle with generating syntactically and semantically correct code for XMI instance models. 
This limitation arises from the inherent complexity of XMI, which involves deeply nested structures and numerous XML tags, as well as the limited availability of training data tailored to this specific domain.
To address this, we use the \ac{LLM} to generate a structured, intermediate textual representation of the instance model, which we refer to as the conceptual instance model.
This representation captures all necessary structural information but avoids the verbosity and strict formatting constraints of XMI.

We further implement an instance compiler, which takes as input the original Ecore metamodel file and the conceptual instance model, and generates a valid XMI instance model file.
The output is guaranteed to conform to the required syntax of the metamodel and the structure of \ac{XMI} format, enabling seamless use in model-based tools.
In addition, the instance compiler performs validation by identifying any unmatched elements between the conceptual instance model and the metamodel. 
Based on these logs, human engineers can inspect and refine the generated XMI model to ensure completeness and correctness.

\subsection{Few-shot Prompting}
Prompting is the primary means of interacting with any \ac{LLM}-based system.
The effectiveness of a prompt can vary not only across different models but even across different versions of the same model. 
As state-of-the-art LLMs continue to improve rapidly, the performance gap between different prompting strategies may become less significant over time.
In this paper, our goal is not to identify the optimal prompt for conceptual instance model generation tailored to a specific \ac{LLM}. 
Instead, we aim to propose a workflow that supports the general task of instance model generation. 
To that end, we adopt an intuitive few-shot prompting strategy.

The complete prompt used in our method is shown in Listing~\ref{lst:prompt}.
The \ac{LLM} input consists of a system prompt and a user prompt. 
The system prompt provides a general description of the task, generation rules, a structural template, and examples relevant to conceptual instance model generation. The structure and template for the conceptual instance model are further discussed in Subsection~\ref{sec:conceptual}.
As few-shot examples, we include a set of inputs comprising the PlantUML-based metamodel information, a scenario specification, and a corresponding reference conceptual instance model.
In our experiments, we provide two examples within the system prompt, due to the limited availability of test data.
The user prompt contains the actual metamodel information and specifications for the system to be modeled. Based on this input, the \ac{LLM} generates the corresponding conceptual instance model.

\begin{lstlisting}[caption={Few-shot prompting for conceptual instance model generation},label={lst:prompt},breaklines=true]
# System:
You are given a meta-model, which defines ONLY the allowed classes, attributes, and associations. Additionally, you are provided with a scenario description, which explicitly specifies the valid instances and relationships. Your task is to generate a Conceptual Instance Model by structuring the information from the scenario description strictly according to the provided meta-model and Json template.

### STRICT RULES
- DO NOT infer missing details. 
- ONLY include what is explicitly provided in the scenario.
- ONLY include what is explicitly provided in the meta model.
- DO NOT create objects from abstract class.

### Output Json Format - Conceptual Instance Model (STRICT TEMPLATE)
{{Template of conceptual instance model}}

### Few-shot Examples
You are provided with the following few-shot examples to help you understand the task.

#### Example
Meta-model information:
{{textual meta-model information}}

Scenario description:
{{textual specifications}}

Generated Conceptual Instance Model:
{{conceptual instance model example}}

# User:
#### Please generate the Conceptual Instance Model follow the template
Meta-model information:
{{Meta-model}}

Scenario description:
{{Scenario description}}

Generated Conceptual Instance Model:
\end{lstlisting}

\subsection{Conceptual Instance Model}\label{sec:conceptual}
We propose a JavaScript Object Notation (JSON)-based conceptual instance model (Listing~\ref{lst:conceputalInstanceModel}) as a simplified and structured representation of the instance model.
Instead of providing a formal JSON schema, we directly define a JSON template to make the LLM-based generation task as straightforward as possible. Generating JSON based on a strict schema would complicate the task and closely resemble metamodel-based instance model generation, which we aim to simplify.
An alternative format for the conceptual instance model could be the PlantUML notations for object diagram.
However, since the object diagram syntax in PlantUML is independent of the class diagram specification, additional mechanisms would be required to explicitly link object instances to their corresponding class definitions. 
Therefore, we adopt the JSON format for its clarity, structure, and ease of parsing.

The conceptual instance model includes all the elements required for object, attribute, and association creation in an XMI instance model, but remove metamodel-level details such as package registration. It also adopts a flatter, less nested structure to reduce complexity.
In the conceptual instance model, each instance object is defined with an unique \textit{InstanceID}, which is later used to specify relationships between objects.
The \textit{type} field specifies the class information of the object, corresponding to the provided metamodel.
Each object has a list of \textit{attributes} and a list \textit{associations}. 
In the \textit{attributes} list, each attribute entry includes the \textit{DataType}, \textit{AttributeName} and \textit{Value}.
For \textit{associations}, we further divide them into two types: \textit{compositions} and \textit{references}.
\textit{Compositions} describe containment relationships between objects. In XMI, composed objects are typically represented in a nested format, where the child object is defined within the parent. This structure poses a challenge for current LLMs to generate reliably. To simplify this, we represent compositions in the conceptual instance model as flat associations using \textit{InstanceID} references.
\textit{References} define non-containment associations to other objects, also using \textit{InstanceID}.

\begin{lstlisting}[caption={Template of conceptual instance model},label={lst:conceputalInstanceModel}]
{
 "<InstanceID>": {
  "type": "<ClassName>",
  "attributes": [
   {
    "dataType": "<DataType>",
    "attributeName": "<AttributeName>",
    "value": "<Value>"
   }
  ],
  "associations": {
   "compositions": [
    {
     "associationName": "<AssociationName>",
     "associatedClassName": "<ClassName>",
     "instanceID": "<InstanceID>"
    }
   ],
   "references": [
    {
     "associationName": "<AssociationName>",
     "associatedClassName": "<ClassName>",
     "instanceID": "<InstanceID>"
    }
   ]
  }
 }
}
\end{lstlisting}

\subsection{Instance Compiler}\label{sec:compiler}
The conceptual instance model provides a formally structured representation of the system, allowing it to be processed and analyzed automatically by computer programs. However, to enable seamless integration into existing model-based development environments and toolchains, it must be transformed into a standard model file format, e.g., XMI.

\begin{algorithm}[t]
	\caption{Instance Compiler Algorithm}
	\label{alg:compiler}
	\begin{algorithmic}[1]
		
		\REQUIRE \textit{metaModel}, \textit{conceptualInstanceModel}
		\ENSURE \textit{XMIFile}
		
		// Load the metamodel
		\STATE \textit{ecoreResource} $\leftarrow$ parse(\textit{metamodel})
		\STATE \textit{metaModelRoot} $\leftarrow$ \textit{ecoreResource}.rootPackage
		\STATE register \textit{metamodelRoot} in the metamodel registry
		
		// Parse the conceptual instance file
		\STATE \textit{fileData} $\leftarrow$ read(\textit{conceptualInstanceFile})
		\STATE \textit{instancesMap} $\leftarrow$ parseJson(\textit{fileData})
		
		// Instantiate objects
		\FOR{(\textit{instanceId}, \textit{instanceData}) \textbf{in} \textit{instancesMap}}
		\STATE \textit{className} $\leftarrow$ \textit{instanceData}.type
		\STATE \textit{eClass} $\leftarrow$ \textit{metamodelRoot}.getEClassifier(\textit{className})
		\IF{\textit{eClass} $\neq$ null \textbf{and} not \textit{eClass}.isAbstract}
		\STATE \textit{objectsMap}[\textit{instanceId}] $\leftarrow$ \textbf{new} eClass()
		\ENDIF
		\ENDFOR
		
		// Populate attributes and associations
		\FOR{(\textit{instanceId}, \textit{instanceData}) \textbf{in} \textit{instancesMap}}
		\STATE \textit{owner} $\leftarrow$ \textit{objectsMap}[\textit{instanceId}]
		\IF{\textit{owner} = null}
		\STATE \textbf{continue}
		\ENDIF
		
		\FOR{\textit{attr} \textbf{in} \textit{instanceData}.attributes}
		\STATE setAttribute(\textit{owner}, \textit{attr})
		\ENDFOR
		
		\FOR{\textit{assoc} \textbf{in} \textit{instanceData}.associations}
		\STATE \textit{targetObj} $\leftarrow$ \textit{objectsMap}[\textit{assoc}.targetId]
		\STATE setAssociation(\textit{owner}, \textit{assoc}.name, \textit{targetObj})
		\ENDFOR
		\ENDFOR
		
		// Export the model to XMI
		\STATE \textit{xmiResource} $\leftarrow$ createXMIResource()
		\FOR{\textit{obj} \textbf{in} \textit{objectsMap}}
		\IF{\textit{obj.eContainer()} = null} 
		\STATE \textit{xmiResource}.append(obj)
		\ENDIF
		\ENDFOR
		\STATE \textit{xmiResource}.save(\textit{XMIFile})
		
	\end{algorithmic}
\end{algorithm}

We design an instance compiler (Algorithm~\ref{alg:compiler}) to convert the conceptual instance model into a valid XMI instance model. The compiler relies on the Python package PyEcore~\cite{Pyecore}, which provides utilities for analyzing Ecore metamodels and generating instance models in XMI format.
The compilation process begins by loading and parsing both the conceptual instance model and the corresponding metamodel. 

The instantiation procedure consists of two main steps: object instantiation, attribute/association assignment.
In the first step, all instance objects are instantiated.
All instance objects defined in the conceptual model are instantiated. For each object, the specified class type is validated against the metamodel. Only classes that exist and are not abstract are used to create objects. Successfully instantiated objects are stored in a map for later reference.
After all objects are created, their attributes and associations are populated. The functions \textit{setAtrribute} and \textit{setAssociation} include checks for type correctness and the existence of the specified attributes or references in the metamodel.

Once all objects and their properties have been defined, the complete model is exported to an XMI file using PyEcore. PyEcore ensures the grammatical correctness of the generated XMI file and searches for and inserts metamodel information, including the package URI, which is often problematic when generating XMI format directly from LLMs. The final XMI model can therefore be directly loaded into standard modeling tools such as Eclipse.

During the compilation process, invalid inputs or inconsistencies in the conceptual instance model are detected either by exceptions raised by PyEcore functions or by validation checks embedded in functions \textit{setAtrribute} and \textit{setAssociation}.
The logs should be reviewed by domain experts to refine and finalize the generated instance model.

\section{Evaluation} \label{sec:evaluation}
To demonstrate the feasibility of the proposed approach, we selected several flagship \acp{LLM}, including the commercial models GPT-4o and o1-preview, as well as the open-source model Llama 3.1 in both 8B and 70B parameter sizes. We conducted experiments to assess the performance of these models within the proposed framework.
The GPT models were accessed via OpenAI's chat completions API.
The Llama 3.1 models (4-bit quantized) were downloaded and deployed locally using Ollama~\cite{Ollama}.

We consider the instance model generation a deterministic task since the output must strictly follow the syntax defined by the XMI format and the Ecore metamodel, and semantics inferred from the given scenario description.  
Accordingly, we set the temperature of GPT-4o and Llama 3.1 models to 0 to ensure deterministic outputs.
However, for GPT o1-preview, due to its inherent self-reflective reasoning mechanism, the temperature is fixed at 1 and cannot be modified.

As a baseline, we first evaluated all selected LLMs using an intuitive generation approach, where each LLM was tasked with directly generating an XMI instance model. The prompts used in this approach were adapted from Listing~\ref{lst:prompt}, removing the template for conceptual instance model and using actual XMI code as few shot examples. We compared the validity of instance models produced by this intuitive method against those generated using the proposed approach.
Following this comparison, we further analyzed the semantic quality of the models generated using the proposed method.

It is important to note that our goal is not to benchmark individual LLMs, as their capabilities are evolving rapidly. Instead, our focus is on showcasing how the proposed method benefits the instance model generation process when applied to both commercial and open-source LLMs.

\subsection{Test Set Collection}
As no existing dataset supports the task of generating XMI instance models from Ecore metamodels and natural language specifications, we manually constructed a test set. This dataset consists of 26 Ecore metamodels selected from the OCL dataset in~\cite{Pan2024}, which have clear real-world semantics

For each metamodel, corresponding natural language specifications were generated using ChatGPT and refined by human experts. Manually-created XMI instance models were also prepared, resulting in 26 pairs of metamodels, natural language descriptions, and their corresponding instance models. In our experiements, two of these pairs were used as few-shot examples. The remaining 24 were used for evaluation. We make this test set publicly available on Huggingface\footnote{https://huggingface.co/datasets/fpan/text-to-xmi-from-ecore}.

\begin{figure}[t]
	\centering
	\includegraphics[width=\linewidth]{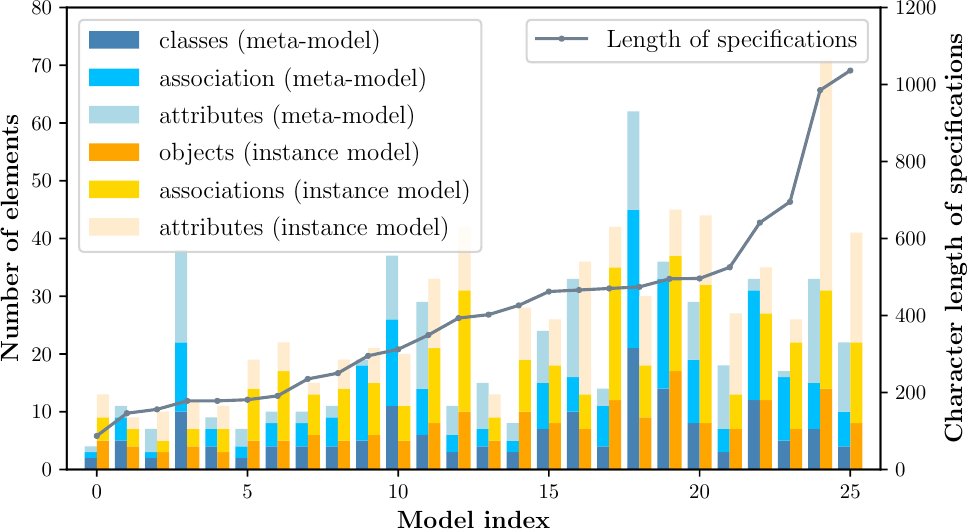}
	\caption{Test set information: Distribution of elements in meta-models and instance models, along with the character length of specifications for each instance model.}
	\label{fig:testset}
\end{figure}

Figure~\ref{fig:testset} provides an overview of the test set. The bar chart shows the distribution of model elements in the metamodels and instance models. The line plot represents the character length of each natural language specification.

\subsection{Evaluation Metrics}
To assess the quality of instance model generation, we designed evaluation metrics from both grammatical and semantic perspectives. Among them, the validity rate is used to assess the overall performance across all generation tasks. The grammatical accuracy, semantic precision, semantic recall, and semantic accuracy are applied to evaluate individual instance models in each generation. We use the average values of these metrics for the final evaluation.

As notation, we define \( E_{gt} \) as the set of elements in the ground truth model and \( E_{g} \) as the set of elements in the generated model. These elements include objects, attributes, and associations.

\subsubsection{Validity Rate (VR)}  
VR measures the proportion of generated instance models that are syntactically correct both in terms of XMI formatting and compliance with the syntax defined by the metamodel. A valid instance model should be loadable and viewable in modeling tools such as Eclipse.

\begin{equation}
	VR = \frac{\text{Grammatically Correct Models}}{\text{Total Generated Models}}
\end{equation}

\subsubsection{Grammatical Accuracy (GA)}  
GA represents the proportion of elements in a conceptual instance model that conform to the syntax rules defined by the metamodel and can be successfully compiled into valid XMI.

\begin{equation}
	GA = \frac{\text{Grammatically Correct Elements}}{\text{Total Generated Elements}}
\end{equation}

\subsubsection{Semantic Precision (SP)}  
SP indicates the percentage of elements in a generated model that are precise and relevant to the natural language specifications. A higher SP indicates that fewer hallucinations were generated.  

\begin{equation}
	SP = \frac{|E_g \cap E_{gt}|}{|E_g|}
\end{equation}

\subsubsection{Semantic Recall (SR)}  
SR measures how many of the required elements from the specification were correctly generated in a generated instance model. A higher SR value signifies that fewer required elements are missing.

\begin{equation}
	SR = \frac{|E_g \cap E_{gt}|}{|E_{gt}|}
\end{equation}

\subsubsection{Semantic Accuracy (SA)}  
SA assesses the overall semantic correctness of a generated model by comparing the semantically correct elements to the complete set of relevant elements, including both missing and hallucinated ones

\begin{equation}
	SA = \frac{|E_g \cap E_{gt}|}{|E_g \cup E_{gt}|}
\end{equation}

\subsection{Experiment Results}
We present the experimental results in two comparative groups: one focusing on the overall validity of generated instance models using different methods, and the other providing a more detailed analysis of model content and semantics

The first comparison evaluates the VR of instance models generated using both the intuitive generation approach and the proposed method.
As shown in Table~\ref{tab:successful_rate}, the intuitive generation method produced a low validity rate. Most outputs contained syntax errors and were unusable in modeling tools. A common issue was incorrect or missing XMI schema definitions.

To partially mitigate this, we manually fixed schema-related issues and re-evaluated the outputs. After repairs, models generated by GPT-4o, o1-preview, and Llama 3.1-70B showed marked improvements, with around half being grammatically correct. However, Llama 3.1-8B still had a very low success rate (4\%).
This low validity significantly hinders further evaluation, especially regarding semantic quality.

In contrast, the proposed method ensures syntactic correctness of the generated model by offloading XMI complexity to a dedicated instance compiler, allowing the LLM to focus on generating semantically meaningful conceptual instance models.
The overall grammatical accuracy of these conceptual models is promising, with even Llama 3.1-8B achieving a GA score of 88\% (see Table~\ref{table: comparison}). A detailed analysis of the GA of the conceptual instance models, along with the SP, SR, and SA scores of the generated XMI models using the proposed approach, is presented in the following part.

\begin{table}[h]
	\vspace{0.5cm}
	\caption{Comparison of validity rates for XMI instance models generated using intuitive and proposed methods. The validity rate represents the usability of the generated XMI instance model.}
	\centering
	\label{tab:successful_rate}
	\sisetup{group-digits=false,table-number-alignment=right,table-parse-only=true}%
	\begin{tabular}{l>{\centering\arraybackslash}m{1.2cm} c>{\centering\arraybackslash}m{1.2cm}}
		\toprule
		LLM & intuitive generation & repaired XMI & proposed method\\
		\midrule
		GPT-4o & 4\% &59\% & 100\% \\
		o1-preview & 8\%  & 46\% & 100\% \\
		Llama 3.1-8B & 0\% & 4\% & 100\%\\
		Llama 3.1-70B & 20 \%  & 46\% & 100\%\\
		\bottomrule
	\end{tabular}
\end{table}

The second group of experiments focuses on the content of the generated models produced using the proposed method with different \acp{LLM}. We collected the average values of evaluation metrics, including  grammatical accuracy, semantic precision, semantic recall, and semantic accuracy, across  execution runs with varying metamodels and specification inputs.

In terms of grammatical accuracy of the conceptual instance models, all \acp{LLM} achieved promising results. Notably, GPT-4o achieved an average GA of 99\%, meaning that nearly all elements in its conceptual instance models could be successfully parsed into valid XMI. The other models also performed well, with o1-preview, Llama 3.1-8B, and Llama 3.1-70B achieving GA scores of 96\%, 88\%, and 90\%, respectively, indicating that the generated elements mostly conformed to the metamodel syntax.

In terms of semantic quality of the final generated instance models, GPT-4o again outperformed the others, achieving the highest average scores in semantic precision, recall, and accuracy. This indicates that its generated models contained fewer hallucinated elements and closely followed the input specifications. The o1-preview and Llama 3.1-70B also demonstrated solid performance overall.
Interestingly, Llama 3.1-70B, an open-source \ac{LLM} with much less parameters compared  to GPT models, achieved a high average semantic recall, comparable to that of GPT-4o, suggesting it is capable of generating most of the required elements from the specifications. However, its relatively lower semantic precision indicates that it tends to hallucinate additional, unspecified elements. 
Llama 3.1-8B showed the weakest performance, with average SP, SR, and SA scores of 74\%, 74\%, and 60\%, respectively. This result was expected given its smaller parameter size. When analyzing performance across individual element types, we observed that all \acp{LLM} performed better at generating objects compared to attributes and associations.

Overall, these results demonstrate that the proposed approach significantly improves the validity and usability of generated instance models. Moreover, it helps to narrow the performance gap between large-scale commercial \acp{LLM} and smaller, open-source models that can be deployed locally.

\begin{table}[h]
	\vspace{0.5cm}
	\caption{Evaluation of grammatical and semantic metrics for generated instance models using the proposed method.}
	\centering
	\label{table: comparison}
	\sisetup{group-digits=false,table-number-alignment=right,table-parse-only=true}%
	
	\begin{subtable}[t]{0.5\textwidth}
		  \centering
		  \caption{All instance model elements}
		
		\begin{tabular}{lcccc}
			\toprule
			
			LLM & $\overline{GA}\uparrow$ & $\overline{SP}\uparrow$ & $\overline{SR}\uparrow$ & $\overline{SA}\uparrow$  \\
			\midrule
			
			GPT-4o 			& 99\% 	& 95\%	& 96\% 	& 93\%	\\
			o1-preview 	& 96\%	& 91\%	& 92\%	& 87\%	\\
			Llama 3.1-8B 	& 88\% 	& 74\% 	& 74\% 	& 60\%  \\
			Llama 3.1-70B 	& 90\%  & 84\% 	& 95\%  & 81\%	\\
			\bottomrule
		\end{tabular}
	\end{subtable}
	\vspace{0.2cm}
	
		\begin{subtable}[t]{0.5\textwidth}
		\centering
		\caption{Objects}
		
		\begin{tabular}{lcccc}
			\toprule
			
			LLM & $\overline{GA}\uparrow$ & $\overline{SP}\uparrow$ & $\overline{SR}\uparrow$ & $\overline{SA}\uparrow$  \\
			\midrule
			
			GPT-4o 			& 100\%& 99\% & 98\% & 97\%	\\
			o1-preview 	& 96\% & 94\% & 96\% & 92\%	\\
			Llama 3.1-8B 	& 96\% & 81\% & 90\% & 75\%	\\
			Llama 3.1-70B 	& 99\% & 96\% & 99\% & 95\%	\\
			\bottomrule
		\end{tabular}
	\end{subtable}
	\vspace{0.2cm}
	
		\begin{subtable}[t]{0.5\textwidth}
		\centering
		\caption{Associations}
		
		\begin{tabular}{lcccc}
			\toprule
			
			LLM & $\overline{GA}\uparrow$ & $\overline{SP}\uparrow$ & $\overline{SR}\uparrow$ & $\overline{SA}\uparrow$  \\
			\midrule
			
			GPT-4o 			& 100\%& 96\% & 96\% & 94\% 	\\
			o1-preview 	& 94\% & 88\% & 91\% & 85\%	\\
			Llama 3.1-8B 	& 76\% & 76\% & 69\% & 65\%	\\
			Llama 3.1-70B 	& 92\% & 86\% & 91\% & 81\%	\\
			\bottomrule
		\end{tabular}
	\end{subtable}
	\vspace{0.2cm}
	
		\begin{subtable}[t]{0.5\textwidth}
		\centering
		\caption{Attributes}
		
		\begin{tabular}{lcccc}
			\toprule
			
			LLM & $\overline{GA}\uparrow$ & $\overline{SP}\uparrow$ & $\overline{SR}\uparrow$ & $\overline{SA}\uparrow$  \\
			\midrule
			
			GPT-4o 			& 99\% & 93\%  & 97\% & 91\%	\\
			o1-preview 	& 99\% & 93\%  & 93\% & 88\%	\\
			Llama 3.1-8B 	& 84\% & 68\%  & 73\% & 57\%	\\
			Llama 3.1-70B 	& 88\% & 81\%  & 97\% & 81\%	\\
			\bottomrule
		\end{tabular}
	\end{subtable}
	
\end{table}

\section{Conclusion} \label{sec:conclusion}
This work presents a novel approach for instance model generation based on a given metamodel and natural language specifications. Current \acp{LLM} face limitations when directly generating valid XMI instance models. To address this, the proposed method leverages LLMs to produce a structured conceptual instance model, e.g., in JSON format, which has a simplified grammar compared to XMI. A dedicated transformation algorithm is then used to convert this conceptual model into a valid XMI instance model.
This approach guarantees the syntactic validity of the generated instance models and provides feedback on non-parsable components produced by LLMs. Ensuring model validity enables further semantic analysis and practical use in model-driven engineering workflows.

Evaluation was conducted using a small but representative test set, consisting of manually selected metamodels from existing open-source data, machine-generated natural language specifications reviewed by humans, and hand-crafted reference instance models. The results demonstrate that the proposed method significantly improves instance model generation performance, ensuring validity for both commercial and open-source LLMs.
In terms of semantic performance, the large commercial models, GPT-4o and o1-preview, generally outperformed the smaller open-source models, Llama 3.1-8B and Llama 3.1-70B. However, Llama 3.1-70B achieved a semantic recall comparable to GPT-4o, demonstrating the feasibility of applying the proposed method even with locally deployed, smaller-scale open-source LLMs.

As future work, we plan to extend the test set, especially by including larger and more complex metamodels, to enable a more comprehensive evaluation. In addition, we intend to explore the integration of Retrieval-Augmented Generation (RAG) techniques to better handle large metamodel inputs. Investigating iterative generation strategies will also be essential for supporting the creation of large instance models. Lastly, we aim to develop mechanisms for detecting and reducing hallucinations in LLM-generated outputs to further enhance model quality.

\bibliographystyle{IEEEtran}
\balance
\bibliography{ref}

\end{document}